\documentclass{ws-procs975x65}

\begin{document}

\title{COLORON MODELS AND LHC PHENOMENOLOGY}

\author{ELIZABETH H SIMMONS$^*$, ANUPAMA ATRE, R. SEKHAR CHIVUKULA, \\ PAWIN ITTISAMAI and NATASCIA VIGNAROLI }

\address{Department of Physics and Astronomy, Michigan State University\\
East Lansing, MI 48824, USA\\
$^*$E-mail: esimmons@msu.edu\\
www.msu.edu}

\author{ARSHAM FARZINNIA}

\address{Institute of Modern Physics, Center for High Energy Physics Research,\\
Tsinghua University, Beijing 100084, China}

\author{ROSHAN FOADI}

\address{Service de Physique Theorique, Universite Libre de Bruxelles, Brussels, Belgium\\
and Centre for Cosmology, Particle Physics and Cosmology (CP3), \\
Universite catholique de Louvain, B-1348 Louvain-la-Neuve, Belgium}

\begin{abstract}
This talk discusses the possibility of new physics within the strong gauge interactions, specifically the idea of an extended color gauge group that is spontaneously broken to QCD.  After a brief review of the literature, three of our recent pieces of work on coloron phenomenology are summarized.  First, some key results on coloron production to NLO at hadron colliders are described.   Next, a method of using associated production of colorons and weak vector bosons to better determine coloron couplings is discussed.  Finally,  a new model that naturally realizes flavor physics is reviewed.
\end{abstract}

\keywords{BSM physics, colorons, LHC phenomenology}

\bodymatter

\section{Introduction}\label{sec:intro}
The LHC Era is proving to be extremely exciting: not only have the experiments decisively rediscovered all of the familiar particles of the Standard Model, confirming that the operations of the accelerator and detectors are well understood, but both ATLAS \cite{:2012gk} and CMS \cite{:2012gu} have also found a new scalar particle that appears to be the long-awaited Higgs Boson. Since the Standard Model with one Higgs doublet is not natural up to arbitrarily high energies, and since it leaves many questions (including the origin of flavor) unanswered, we anticipate that some physics beyond the Standard Model remains to be discovered. 

An intriguing possibility is that an extended color gauge sector may exist.  In particular, new colored states beyond the familiar quarks and gluons could be awaiting discovery at the LHC.  These could reflect a variety of kinds of theories beyond the standard model.  One class of theories are those in which the strong interactions are extended from the standard $SU(3)_{QCD}$ to a larger $SU(3)_1 \times SU(3)_2$ group and in which spontaneous symmetry breaking reduces the larger group to its diagonal subgroup which is identified with $SU(3)_{QCD}$.  These models include topcolor \cite{Hill:1991at}, the flavor-universal topcolor \cite{Chivukula:1996yr}, classic chiral color \cite{Frampton:1987dn}, chiral color with unequal gauge couplings \cite{Martynov:2009en} and a newer flavor non-universal chiral color model \cite{Frampton:2009rk}. Each of these models includes new heavy colored gauge bosons (colorons, topgluons, or axigluons) transforming as a color octet.  Other theories with new color octet states include theories of new extra spacetime dimensions that incorporate Kaluza-Klein partners for the gluons, as in Refs.  \cite{Dicus:2000hm, Davoudiasl:2000wi, Lillie:2007yh} or technicolor models with colored technifermions that bind into color-octet techni-rho mesons \cite{Farhi:1980xs}.  An entire catalog of possible new colored states including color sextet fermions, colored scalars, and low-scale string resonances \cite{Antoniadis:1990ew} has also been reviewed in \cite{Han:2010rf}.

If an extended color gauge sector does exist, then there are indications that it could be likely to couple more strongly to the third generation than the light quarks.  For instance, if the new scalar state with a mass of 125 GeV is actually a composite, rather than a fundamental, scalar, it could potentially be a bound state of top quarks\cite{Miransky:1988xi, Miransky:1989ds, Nambu:1989jt, Marciano:1989xd, Bardeen:1989ds}, as realized in topcolor \cite{Hill:1991at}, topcolor-assisted technicolor \cite{Hill:1994hp} and top seesaw \cite{Dobrescu:1997nm, Chivukula:1998wd, He:2001fz} models, and as analyzed phenomenologically in \cite{Chivukula:2012cp, Barbieri:2012tu}. There is also the puzzling question of how to explain the forward-backward asymmetry observed by the Tevatron experiments \cite{Aaltonen:2011kc, Abazov:2011rq} in the production of top-quark pairs.  Models involving flavor non-universal axigluons \cite{Frampton:2009rk} have been cited as a possible explanation, and there has been discussion in the literature \cite{Chivukula:2010fk, Bai:2011ed} of the degree to which the properties of those axigluons would be constrained by data on flavor-changing neutral currents.

This talk discusses recent work on several different aspects of coloron theory and phenomenology.  We begin by reviewing the basic gauge structure of coloron models.  We then present results on coloron production at NLO, indicating how this impacts colliders searches for new colored resonances.   In the following section,  we discuss a method for using production of a coloron in association with a $W$ or $Z$ boson to determine the chiral couplings of the coloron.  Finally, we present a new model for a coloron model that incorporates a realistic theory of flavor physics.  Each of these topics is discussed in greater detail elsewhere, and we provide the reader with the appropriate references.

\section{Gauge Structure}\label{sec:gauge}

The coloron models discussed in this talk are simple, renormalizable, models with the gauge structure 
$SU(3)_1 \times SU(3)_2 \times SU(2)_W \times U(1)_Y$. 
We name the $SU(3)_1 \times SU(3)_2$ gauge bosons $A^a_{1\mu}$ and $A^a_{2\mu}$, respectively, and call the 
corresponding gauge couplings $g_1$,  $g_2$. The two $SU(3)$ gauge couplings are related to the QCD coupling $g_S$ through
\begin{equation}
g_S=g_1\sin\omega=g_2\cos\omega ,
\end{equation}
where $\omega$ is a new gauge mixing angle.

Gauge symmetry breaking occurs in two steps:
\begin{itemize}
\item $SU(3)_1 \times SU(3)_2 \to SU(3)_C$ due to the (diagonal) expectation value
$\langle \Phi \rangle \propto u \cdot {\cal I}$, where the fundamental or composite scalar, $\Phi$, transforms as a $(3,\bar{3})$ under
$SU(3)_1 \times SU(3)_2$ and ${\cal I}$ is the identity matrix,

\item $SU(2)_W \times U(1)_Y \to U(1)_{em}$ in the usual way due to a Higgs doublet
$\phi$ transforming as a $2_{1/2}$ of the electroweak group, and with the usual
vacuum expectation value given by $v \approx 246$ GeV.
\end{itemize}
We will assume that the color-group symmetry breaking occurs at a scale large compared
to the weak scale, $u \gg v$.\footnote{The vacuum expectation values for $\phi$ and $\Phi$
occur for a choice of parameters in the most general, renormalizable, potential for these fields, and
the vacuum is unique up to an arbitrary global gauge transformation. We will assume that the additional
physical singlet and color-octet fields in $\Phi$ are heavy, and neglect them in what follows.}

The mass-squared matrix for the colored gauge bosons is given by
\begin{equation}
-\dfrac{1}{2} u^2\begin{pmatrix} g_1^2& -g_1 g_2\\ -g_1 g_2& g_2^2 \end{pmatrix}\, .
\end{equation}
Diagonalizing this matrix reveals mass eigenstates $G^a$ and $C^a$
\begin{eqnarray}
G^a_\mu &=& \sin\omega A^a_{1\mu} + \cos\omega A^a_{2\mu} \\
C^a _\mu &=& \cos\omega A^a_{1\mu} - \sin\omega A^a_{2\mu} \label{eq:coloron}
\end{eqnarray}
with masses
\begin{equation}
M_G = 0 \qquad M_C = u \sqrt{g_1^2 + g_2^2} = \dfrac{g_S\, u}{\sin\omega \cos\omega}
\end{equation}

If we name the color current associated with $SU(3)_i$ by the symbol $J_i^{a\mu}$, then the gluon and coloron, respectively couple to the following currents:
\begin{eqnarray}
g_S J_G^{a\mu} &=& g_S (J_1^{a\mu} + J_2^{a\mu})\\
g_S J_C^{a\mu} &=& g_S (\cot\omega J_1^{a\mu} - \tan\omega J_2^{a\mu})
\label{eq:coloroncurrent}
\end{eqnarray}
From this, we calculate the decay width of the coloron into massless color-triplet fermions to be 
\begin{equation}
\Gamma_C = \dfrac{g_S^2 M_C}{24 \pi} \left(n_1 \cot^2\omega + n_2 \tan^2\omega \right)
\end{equation}
where $n_1$ and $n_2$ correspond to the number of Dirac fermion states charged under $SU(3)_1$ and $SU(3)_2$ 
respectively. Finally, we note that at energy scales well below the coloron mass, coloron exchange  may be approximated by the current-current interaction:
\begin{equation}
-\dfrac{g^2_S}{2 M_C^2} J_C^{a\mu} J_{C\mu}^a\,.
\end{equation}

\section{NLO Coloron Production}
In Ref. \cite{Chivukula:2011ng} we report the first complete calculation of QCD corrections to the production of a massive color-octet vector boson. We treat the coloron as an asymptotic state in our calculations, employing the narrow width approximation. Our next-to-leading-order (NLO) calculation includes both virtual corrections as well as corrections arising from the emission of gluons and light quarks, and we demonstrate the reduction in factorization-scale dependence relative to the leading-order (LO) approximation used in previous hadron collider studies.

The QCD NLO calculation of coloron production reported here differs substantially from the classic computation of the QCD NLO
corrections to Drell-Yan production \cite{Altarelli:1979ub}, because the final state is colored. In particular, Drell-Yan
production involves the coupling of the light quarks to a conserved (or, in the case of $W$- or $Z$-mediated processes,
conserved up to quark masses) current. Hence, in computing the NLO corrections to Drell-Yan processes, the current conservation Ward identity insures a cancellation between the UV divergences arising from virtual
quark wave function and vertex corrections. These cancellations do not occur in the calculation of the NLO corrections to coloron production, because of vertex corrections involving the 3-point non-Abelian colored-boson vertices. We use the ``pinch technique" \cite{Binosi:2009qm} to divide the problematic non-Abelian vertex corrections into two pieces -- a ``pinched" piece whose UV divergence contributes to the renormalization of the coloron wavefunction (and, ultimately, a renormalization of the coloron coupling) and an ``unpinched" part whose UV divergence (when combined with an Abelian vertex correction)
cancels against the UV divergences in quark wavefunction renormalization. Once the UV divergences are properly accounted for, the IR divergences cancel in the usual way: the IR divergences arising from real quark or gluon emission cancel against the IR divergences in the virtual corrections, and the IR divergences arising from collinear quarks or gluons in the initial state are absorbed in the properly defined parton distribution functions (PDFs).

We compute the gauge-, quark-, and self-couplings of the coloron assuming 
the extended color structure described in Sec. \ref{sec:gauge}. The calculation yields
the minimal coupling of gluons to colorons, and
allows for the most general couplings of quarks to colorons.  The cancellation of UV divergences described above, however,
occurs only when the 3-coloron coupling has the strength that arises from the dimension-four gauge-kinetic energy terms of
the extended color gauge structure. Our computation applies directly
to any theory with this structure, {\it i.e.} to massive color-octet vector bosons in axigluon, topcolor, and coloron models.
In general, the triple coupling of KK gluons in extra-dimensional models, or of colored technivector mesons in
technicolor models, will not follow this pattern. However our results apply approximately to these cases
as well, to the extent that the $SU(3)_{1} \times SU(3)_{2}$
model is a good low-energy effective theory for the extra dimensional model (a ``two-site" approximation in
the language of deconstruction \cite{ArkaniHamed:2001ca,Hill:2000mu}) or for the technicolor theory (a hidden local symmetry approximation for the effective technivector meson sector \cite{Bando:1984ej,Bando:1985rf}).

\begin{figure}
\includegraphics[width=2.4in]{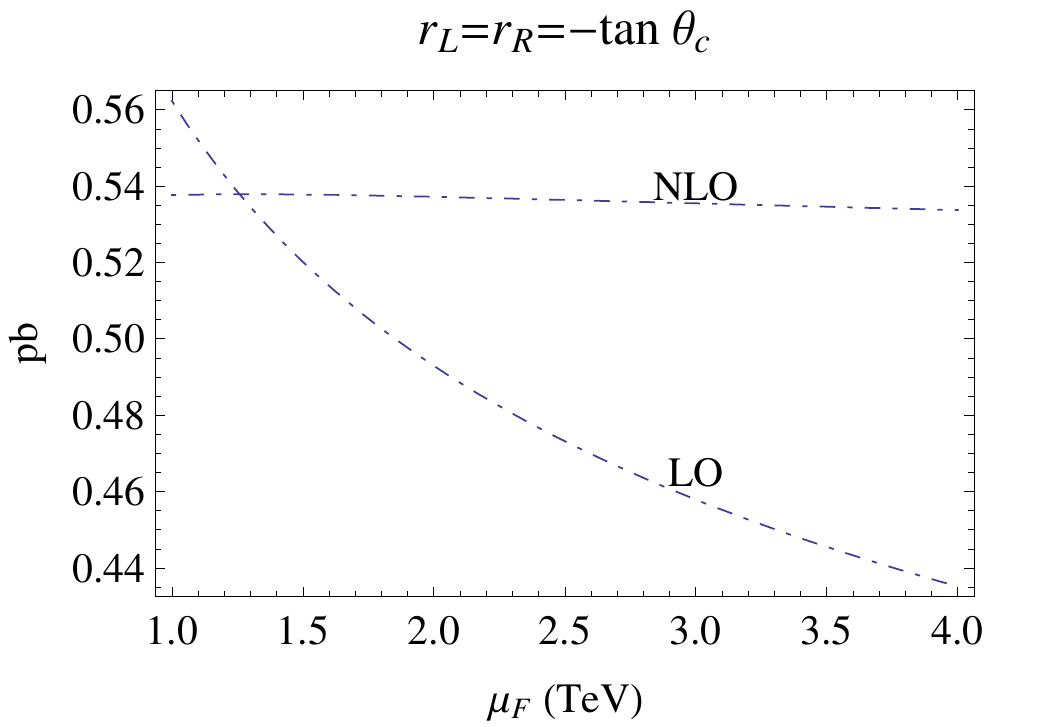}
\includegraphics[width=2.4in]{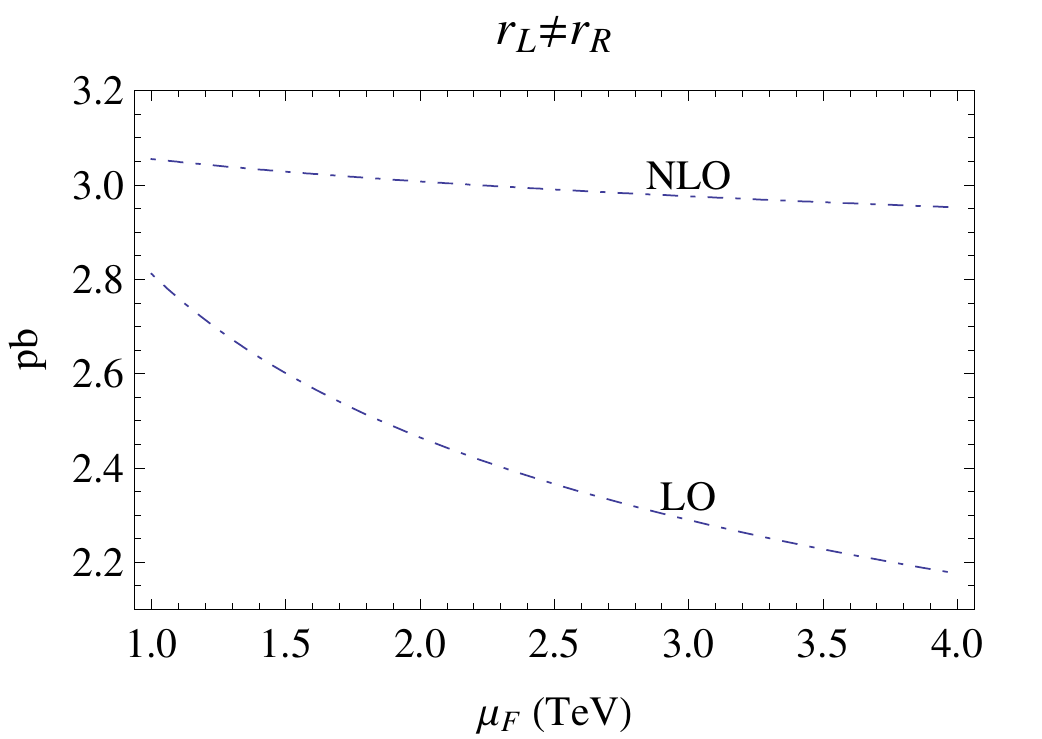}
\caption{Dependence of LO and NLO cross sections at the LHC ($\sqrt{s}=$ 7 TeV), as a function of factorization scale $\mu_F$
for $M_C$ = 2.0 TeV, $\sin^2\theta_c\vert_{\mu = 2.0\,{\rm TeV}} = 0.25$, and two quark charge assignments. The NLO cross section has a much weaker (formally, two-loop) residual scale-dependence.}
\label{fig:scale-dependence}
\end{figure}

\begin{figure}
\includegraphics[width=2.4in]{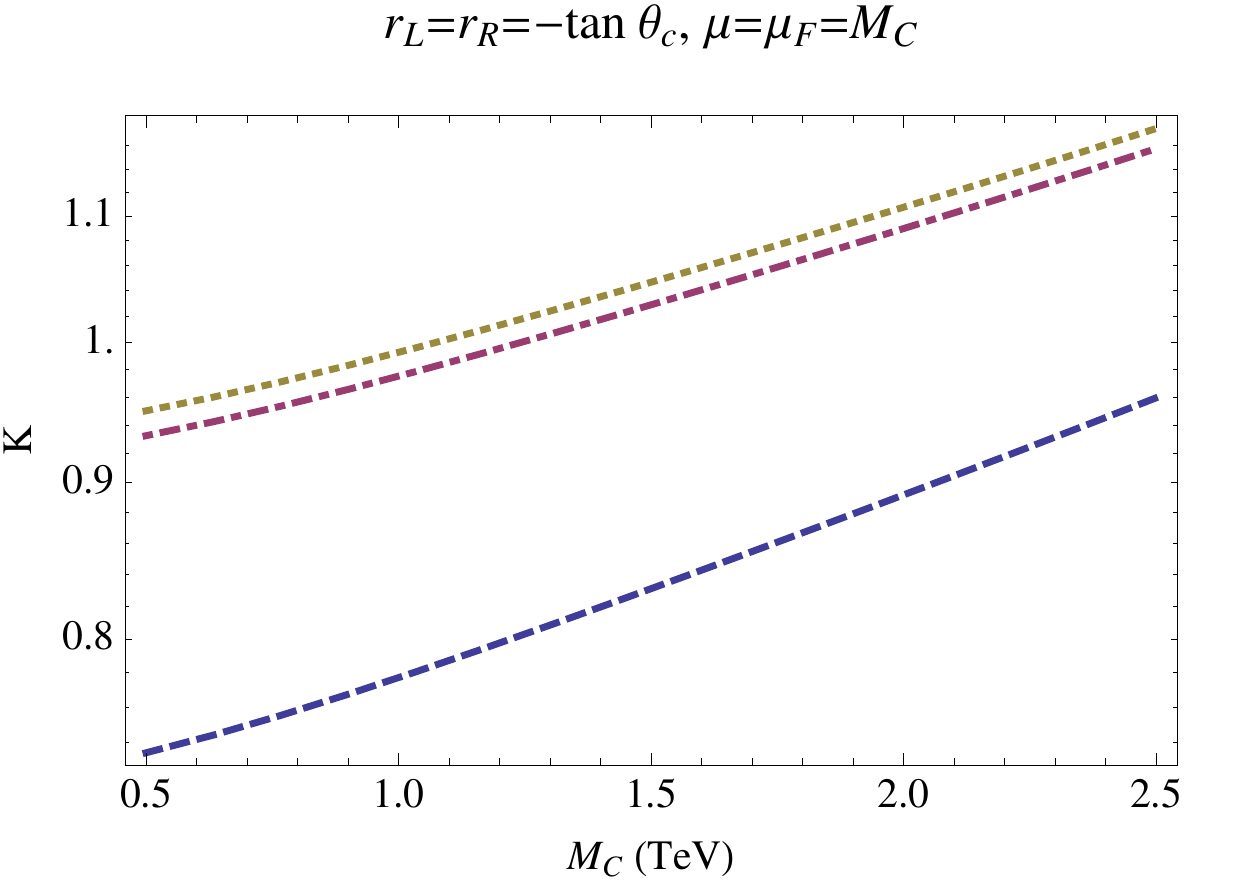}
\includegraphics[width=2.4in]{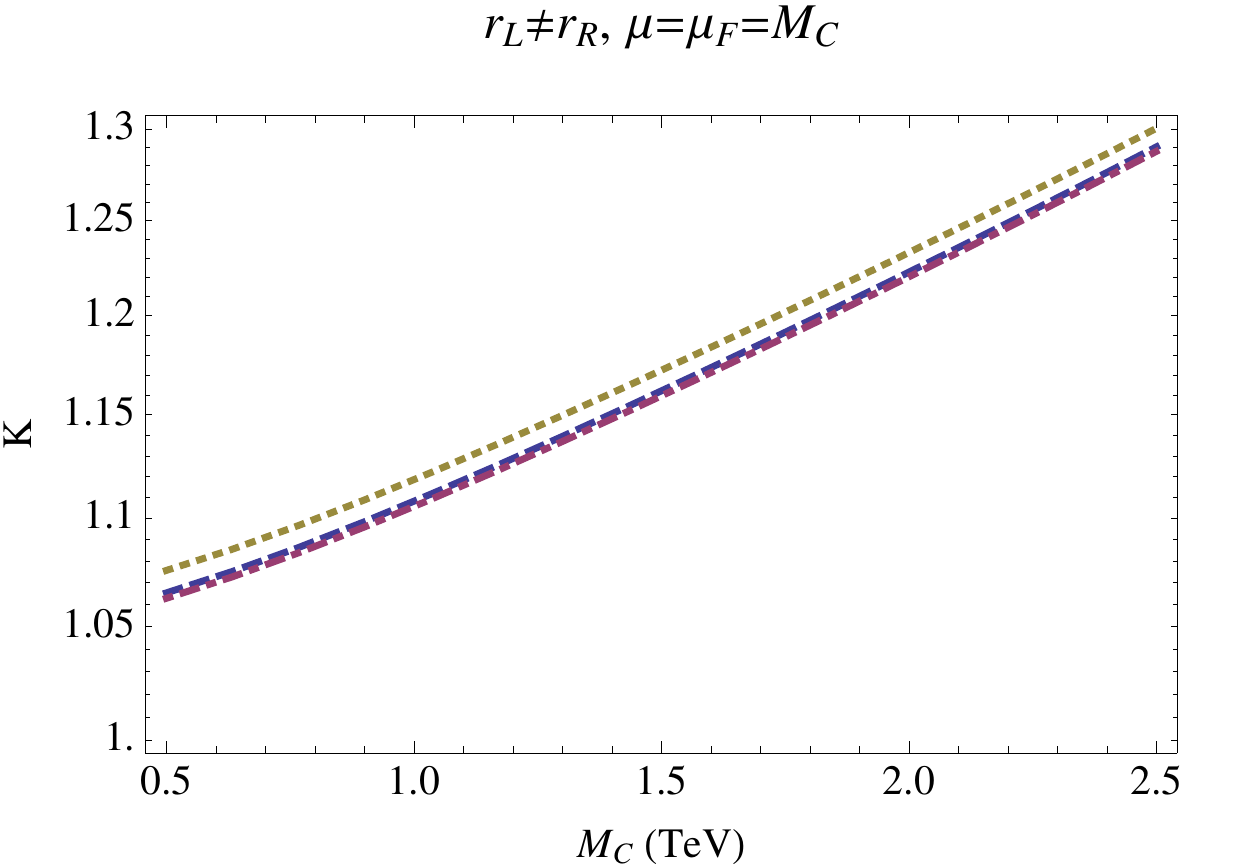}
\caption{``$K$-factor", the ratio of the NLO to LO cross section for coloron production at the LHC ($\sqrt{s}=7$ TeV), plotted as a function of $M_C$ for $\sin^2\theta_c$ = 0.05 (dashed), 0.25 (dot-dashed) and 0.50 (dotted), $\mu_F=M_C$, and the two quark charge assignments.}
\label{fig:Kfactor}
\end{figure}

We find that incorporating NLO corrections makes a substantial difference to the production cross-section and kinematic distributions of the colorons.  Depending on the coloron mass, the production rate can differ significantly from tree-level predictions.  As searches for new resonances decaying to dijets are routinely undertaken by the LHC experiments, taking these results into account is important for setting accurate limits.  Moreover, at NLO, one sees that singly-produced colorons can have substantial transverse momentum (while at tree-level they should have nearly none).

The figures illustrating our results use the following notation:  We denote the gauge boson mixing angle $\omega$ by $\theta_c$ (as was done in Ref. \cite{Chivukula:2011ng}) and we call the coupling of a left(right)-handed quark to colorons, divided by the quark's coupling to gluons $g_S$, by the name $r_L$ ($r_R$).   Plots are shown for two of the three possible flavor-universal charge assignments for the quarks:  all quarks charged under $SU(3)_1$ [$r_L=r_R=-\tan\theta_c$]  and left(right)-handed quarks charged under $SU(3)_{1,(2)}$  [$r_L\neq r_R$]; results for the third case,  in which all quarks charged under $SU(3)_2$ [$r_L=r_R=\cot\theta_c$], are substantially similar to those for the first case.  

Fig.~\ref{fig:scale-dependence} shows that the scale-dependence of the LO cross section is of order 30\% while at NLO  this is reduced to about  2\% percent.  Fig. \ref{fig:Kfactor} shows the ``$K$-factor" for coloron production.
\begin{equation}
K(M_C,\sin\theta_c\vert_{\mu=M_C},\mu_F=M_C) \equiv \frac{\sigma^{NLO}(M_C,\sin\theta_c\vert_{\mu=M_C},\mu_F=M_C)}{\sigma^{LO}(M_C,\sin\theta_c\vert_{\mu=M_C},\mu_F=M_C)}~,
\label{eq:Kfactor}
\end{equation}
for $\sin^2\theta_c$ = 0.05 (dashed), 0.25 (dot-dashed) and 0.50 (dotted).  Again, we see that the NLO corrections are of order 30\%.  We have provided detailed tables of our results for the coloron K-factors for the LHC and Tevatron at a variety of collision energies in Refs. \cite{Chivukula:2011ng, Chivukula:2013xla}.    

Finally, we found that of order 30\% of the colorons in this generic model with masses in the TeV range are produced with $p_T \ge 200$ GeV, once NLO corrections are taken into account.  In other words, the tree-level picture of a new resonance produced at rest and decaying to back-to-back jets receives significant modification at next-to-leading order.  Once a coloron is discovered, studies of the angular distributions of its decay products will need to take this into account.

\section{Associated Coloron Production}

The high production rate of a colored resonance (due to the strong coupling value) and the simple topology of the final state (decay into two jets) makes the search for di-jet resonances one of the early signatures that are studied at hadron colliders. Once a new colored resonance is discovered, measuring its properties will be the next important task. The di-jet invariant mass $m_{jj}$ and the angular distributions of energetic jets relative to the beam axis are sensitive observables to determine the properties, such as mass and spin of the resonance. Although one can constrain the coupling strength of the colored resonance to the Standard Model (SM) quarks using the total cross section, this is not sufficient to determine the chiral structure of the couplings. 

\begin{figure}[tb]
\includegraphics[width=2.3in]{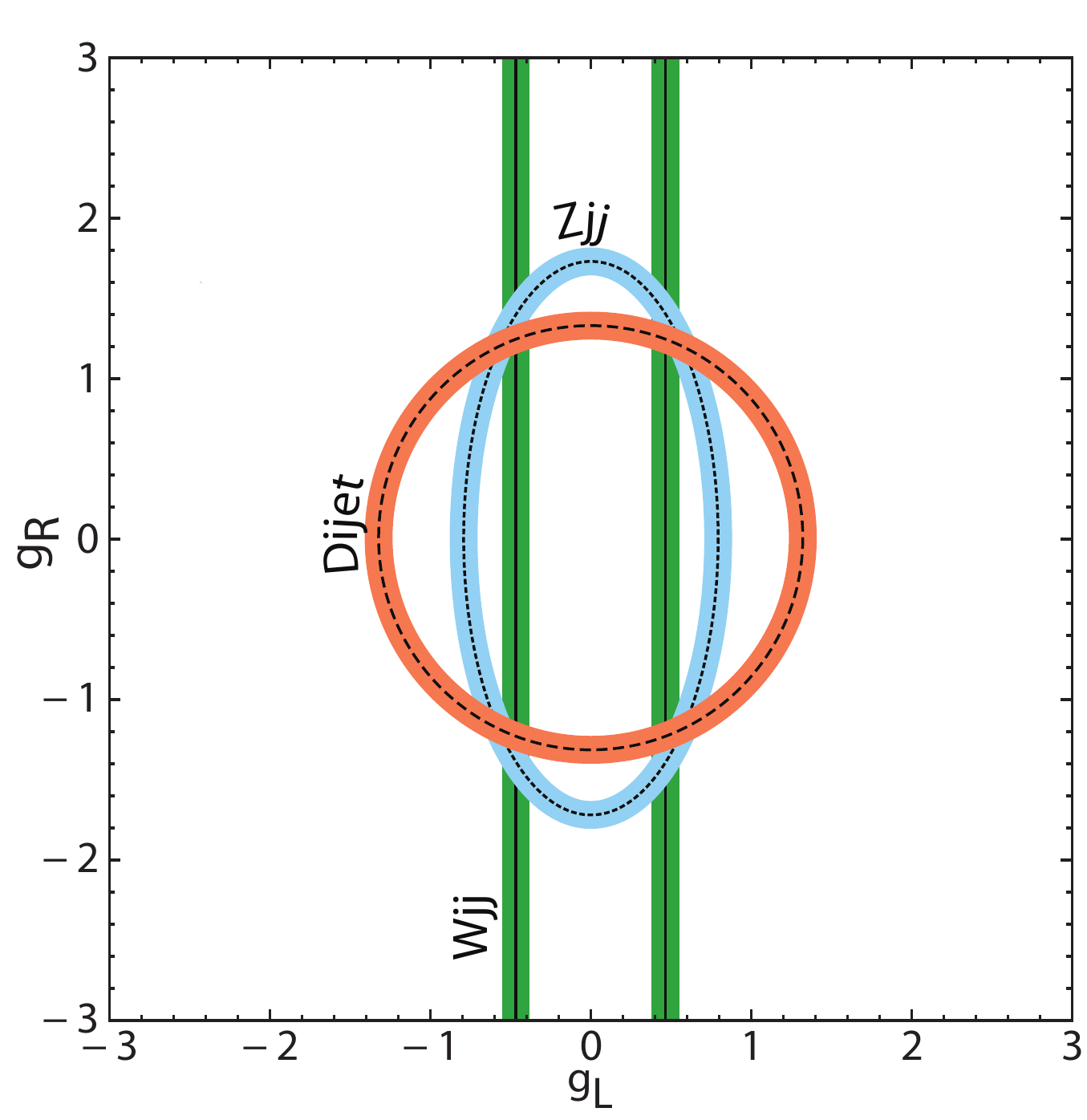}
\includegraphics[width=2.65in]{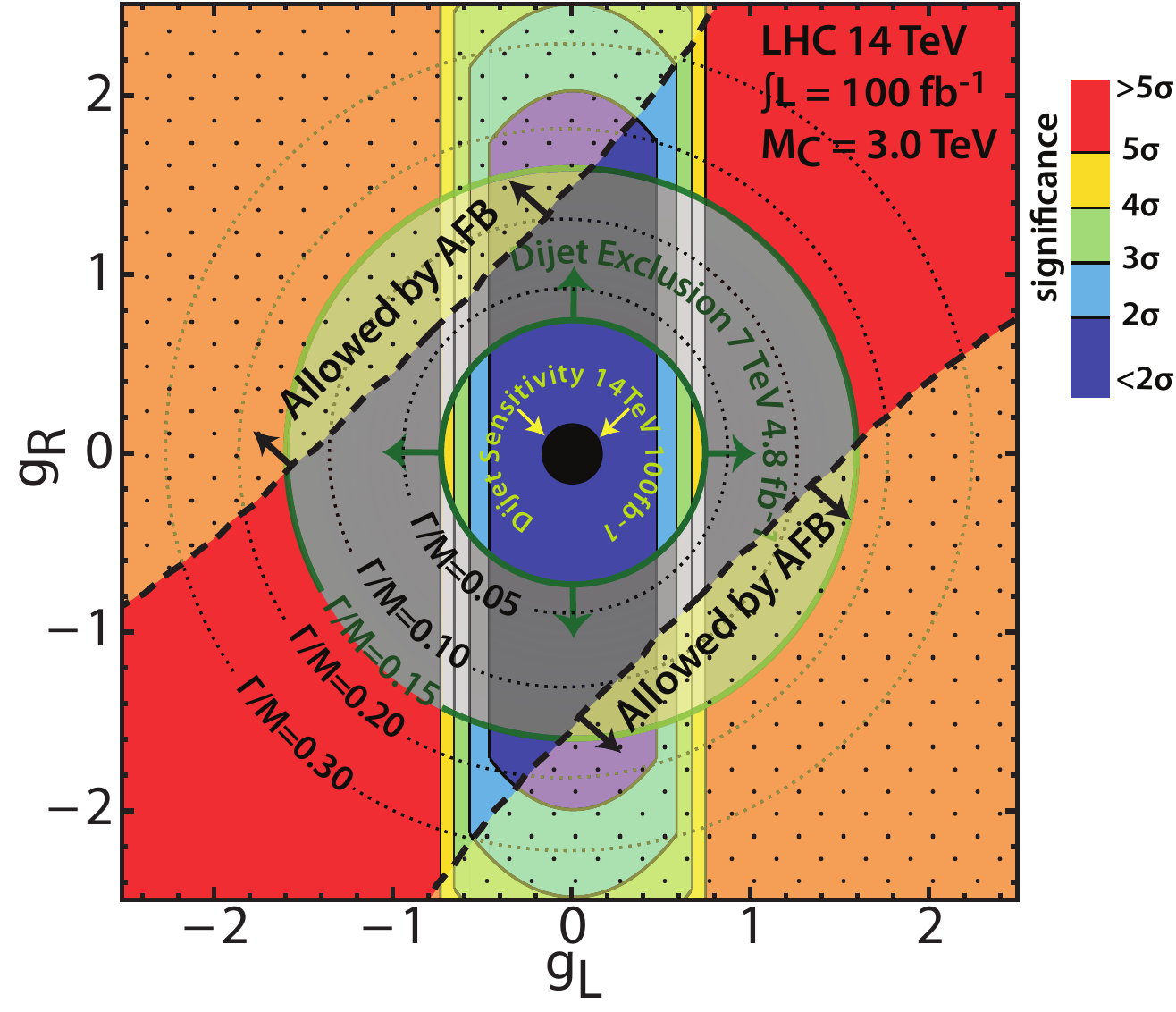}
\caption{ 
Left:  A cartoon illustration of the form of prospective constraints on chiral couplings from the di-jet channel (dashed black circle with red band), the channel with associated production of a $W$ boson (solid black parallel lines with green band) and the channel with associated production of a $Z$ boson (dotted black ellipse with blue band). Combining the constraints from different channels will narrow the range of allowed couplings. Right:  same as Fig.~\ref{fig:cwcz3.5tev}(a) but for $M_C = 3$ TeV and 100 $fb^{-1}$ of data. Here we have combined the sensitivity in the associated production channel from both $W$ and $Z$ bosons. The couplings allowed by the current measurement of $A^{tt}_{FB}$ (see text) are shown by the translucent yellow dotted region.}
\label{fig:cartooncouplings}
\end{figure}

In Ref. \cite{Atre:2012gj}, we propose a new channel for studying coloron couplings: the associated production of a $W$ or $Z$ gauge boson with the color-octet at the LHC. The chiral couplings of the weak gauge bosons to the fermions in the associated production channel provides additional information about the chiral structure of the new strong dynamics. Combining the associated production channel with the di-jet channel makes it possible to extract the chiral couplings of the colored resonance because the cross-sections of each channel have a different dependence on the coloron's couplings to fermions. The functional form of the dependance of these measurements on the chiral couplings in the di-jet channel is $ g^2_L + g^2_R$; in the Wjj channel it is $g^2_L$ and in the Zjj channel it is $ a g^2_L + b g^2_R$. A cartoon illustration of these three measurements along with the di-jet measurement is shown in  the left-hand pane of Fig.~\ref{fig:cartooncouplings}. Notice that while combining the different channels will narrow the allowed range of couplings, there remains an  ambiguity in extracting the sign of the couplings. This method of using the associated production of a weak gauge boson to illuminate the properties of a new resonance was studied earlier in the context of the measurement of $Z^\prime$ couplings \cite{Cvetic:1992qv,delAguila:1993ym}. 

%%%
%%%

The color-octet states ($C$) are produced and decay to two jets via the process 
$p p \stackrel {C} {\longrightarrow} j\ j. $
They can also be produced in association with a weak gauge boson via the processes $p p  \stackrel {C} {\longrightarrow} j\ j\ W^\pm$ and $p p  \stackrel {C} {\longrightarrow} j\ j\ Z$
where $j=u,d,s,c,b$. The associated production modes include $s$ and $t$ channel diagrams with the emission of the gauge bosons in either the initial or final state. The final state channels that we study are
$pp\  \to\  \ell^\pm E_T^{miss}\ 2j,\quad \ell^+\ell^-\ 2j$,
coming from $W^\pm(\to \ell^\pm \nu)$ or $Z (\rightarrow \ell^+ \ell^-)$, respectively and $\ell = e, \mu$. Although the inclusion of the $\tau$ lepton in the final state could increase signal statistics, for simplicity we ignore this experimentally more challenging channel.  Our simulations included the relevant backgrounds
\begin{itemize}
\item $W+$ jets, $Z+$ jets with $W, Z$ leptonic decays;
\item top pair production with fully leptonic, semi-leptonic and hadronic decays  (where some final state particles may be missed or mis-identified);
\item single top production leading to a $W^\pm b\ q$ final state;
\item $W^+W^-, W^\pm Z$ and $ZZ$ with all possible decay combinations leading to the final state of interest.
\end{itemize}
For details on the simulations, including cuts and acceptances, see Ref. \cite{Atre:2012gj}

\begin{figure}[tb]
\includegraphics[width=2.4in]{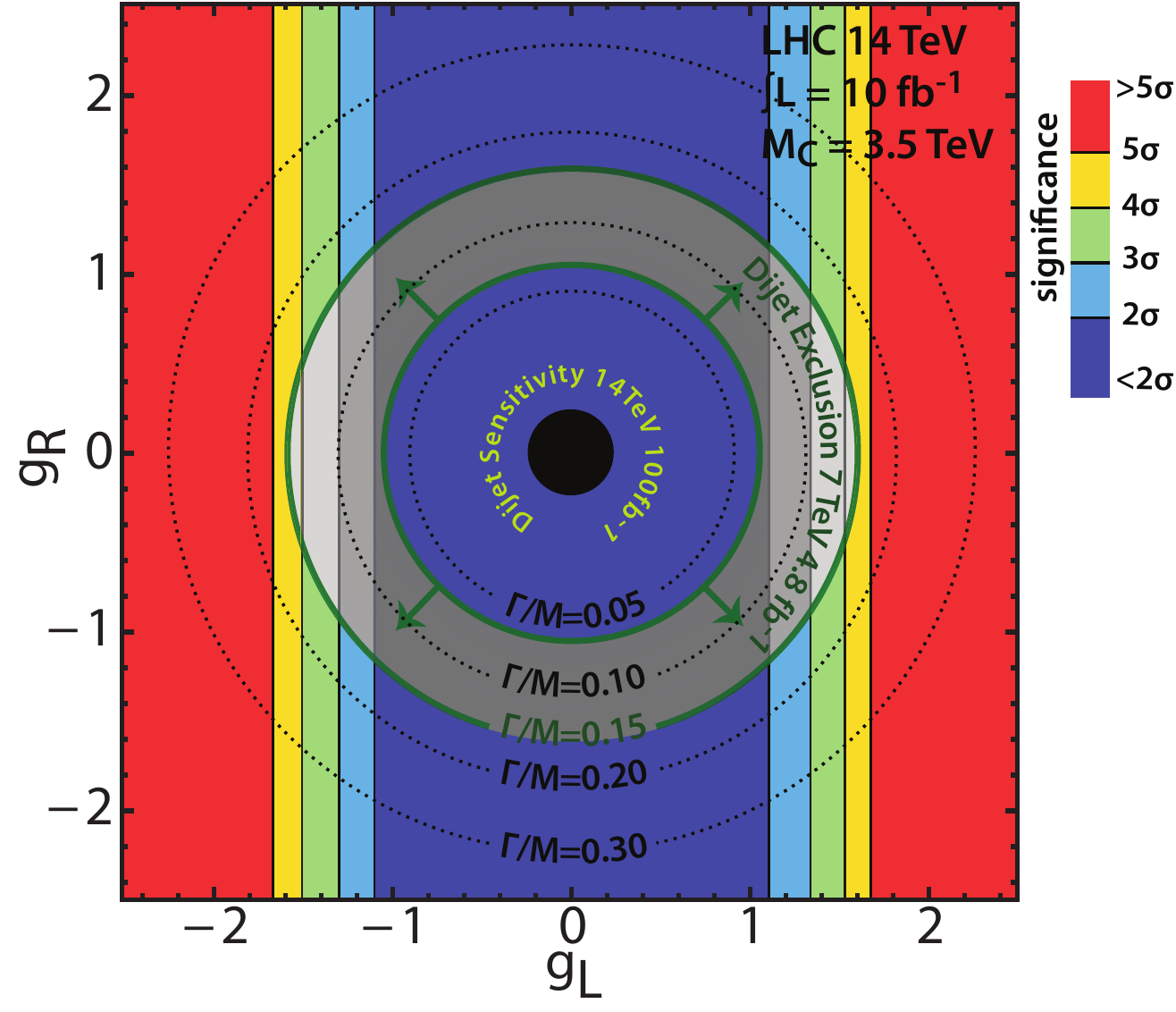}
\includegraphics[width=2.4in]{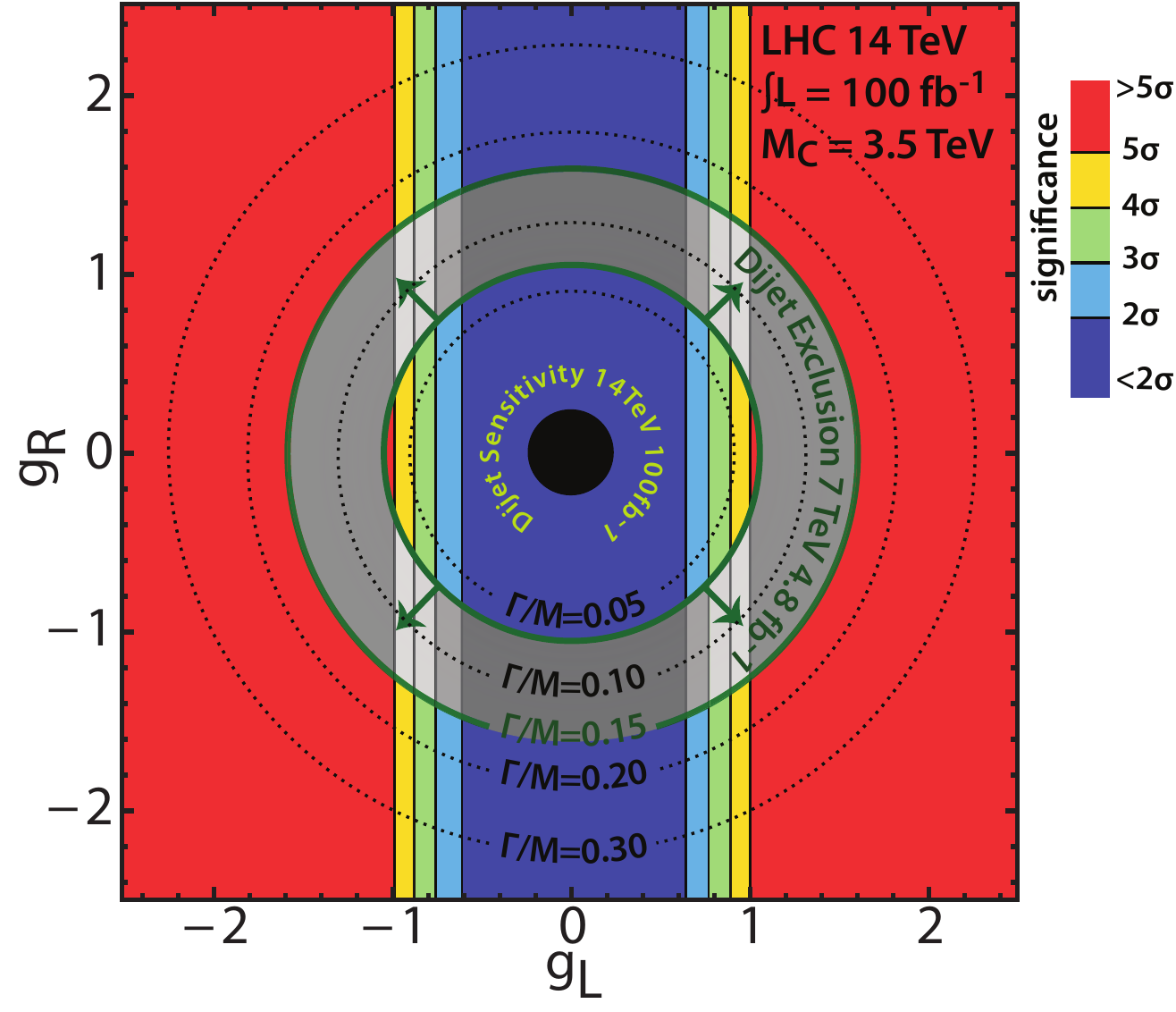}\\
\includegraphics[width=2.4in]{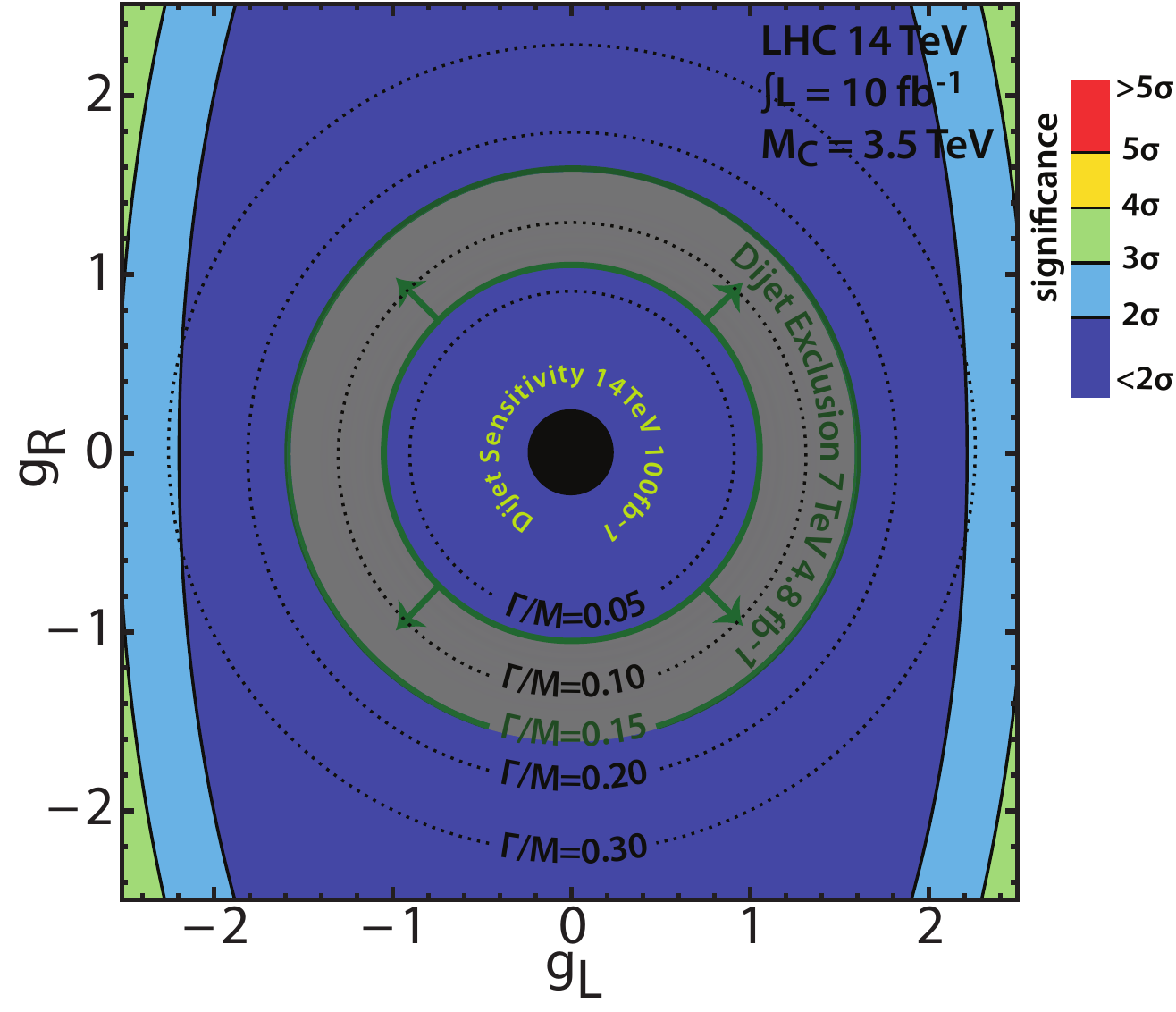}
\includegraphics[width=2.4in]{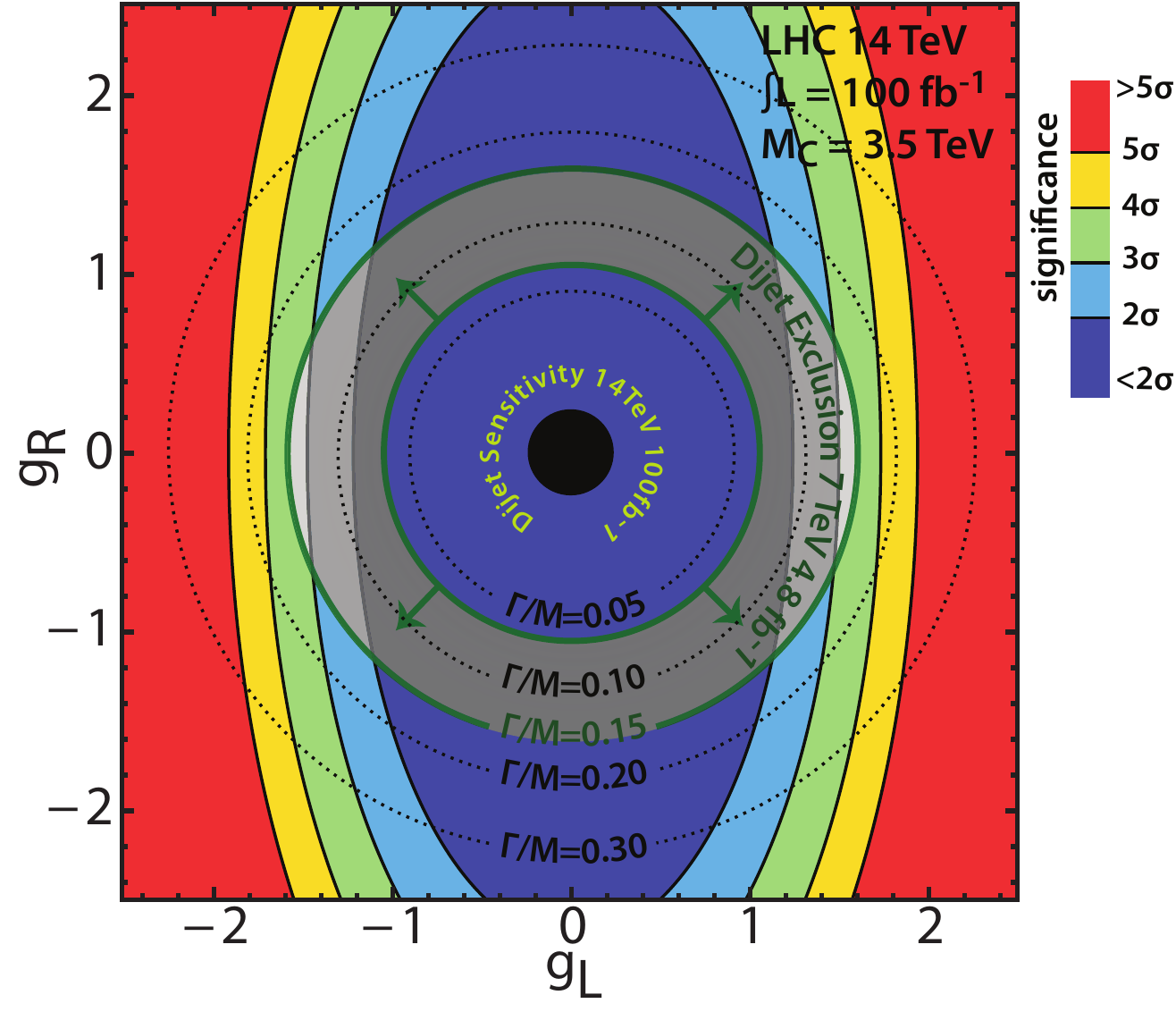}
\caption{ 
(a) Top left: sensitivity plot for a color-octet of mass $M_C = 3.5$ TeV produced in association with a $W$ gauge boson, in the plane of the couplings $g_L^q, g_R^q$ at the LHC with 10 $fb^{-1}$ of data and $\sqrt{s} =  14$ TeV; (b) top right: same for 100 $fb^{-1}$ of data;
  (c) bottom left: same as (a) but for a color-octet produced in association with a $Z$ boson; (d) bottom right: same as (c) but for
  100 $fb^{-1}$ of data. The colored bands show regions with significance from $2\sigma$ to $>5\sigma$. The inner solid green circle is the limit from current direct searches for narrow dijet resonances at the LHC \cite{Aad:2011fq, Aad:2010ae, Chatrchyan:2011ns, Khachatryan:2010jd, atlas:2012-038}. The outer green circle corresponds to a contour with $\Gamma_C/M_C = 0.15$. The faded grey region between the two circles  is excluded for narrow resonances while the region beyond the outer green circle is allowed for broad resonances. The small black region in the center lies beyond the projected dijet sensitivity at the LHC with 100 $fb^{-1}$ of data cite{Gumus:2006mxa}. Black dotted contours indicate the combinations of couplings that give rise to varying widths $\Gamma_C/M_C = 0.05,\  0.10,\  0.20$ and $0.30$. 
 }  
\label{fig:cwcz3.5tev}
\end{figure}

We present the results of our analysis for the LHC in the plane of the couplings $g_L^q, g_R^q$ for different masses, $M_C$, of the color-octet. The sensitivity for the channel with associated production of a $W(Z)$ gauge boson is presented in the upper (lower) panels of Fig.~\ref{fig:cwcz3.5tev}, while the left (right) panels are for integrated luminosity of 10 (100) $fb^{-1}$. The different colored bands represent varying significance of signal observation from $2\sigma$ to greater than $5\sigma$. The black dotted curves are contours of constant widths and we show the curves for several values of $\Gamma_C/M_C = 0.05,\  0.10,\  0.20$ and $0.30$. The small black region in the center lies outside the projected dijet sensitivity (i.e. couplings within the black region result in a dijet production rate too small to be observed) at the LHC with 100 $fb^{-1}$ of data cite{Gumus:2006mxa}. 

The results in Fig.~\ref{fig:cwcz3.5tev} illustrate several features. In the channel with associated production of a $W$ boson, there is no sensitivity in the region near $g_L^q = 0$ due to the left-handed couplings of the $W$ boson, and the sensitivity improves as we move away from the $g_L^q = 0$ axis. The channel with the associated production of a $Z$ boson on the other hand is sensitive to both left and right-handed couplings and sensitivity in the region close to $g_L^q = 0$ is non-zero. The smaller production cross section for this channel along with the small leptonic branching fractions for $Z$ decays limit the gain in sensitivity. Nonetheless this channel provides an additional measurement and hence useful information in untangling the couplings. 

To summarize our results for a variety of coloron masses and couplings with either 10 $fb^{-1}$ or 100 $fb^{-1}$ of data: As expected, the longer run with more data has better sensitivity and can probe masses up to 4.5 (4.0) TeV in the channel with associated production of a $W(Z)$ gauge boson. The projected limit of sensitivity in the dijet channel for the LHC with 100 $fb^{-1}$ data is shown in the figures as the small black circle in the center with small couplings. It is safe to assume that the final acceptance will be at least equal to or even better than the current one and the small black region in the center could shrink even further.  If the LHC were to discover a resonance with such small couplings, one would have to find other novel ways of understanding the chiral structure of couplings as the associated production channel does not have sensitivity in those regions of parameter space. 

Separately, we note that the Tevatron has made a measurement of the top-pair forward-backward asymmetry ($A^{tt}_{FB}$) \cite{cdf_afb5invfb2010, d0_afb4invfb2010}. The authors of Ref.~\cite{Rodrigo:2010gm} have translated this measurement into constraints on the couplings of color-octet resonances. We show this additional constraint for the case of a color-octet with mass $M_C = 3$ TeV in the right-hand pane of Fig.~\ref{fig:cartooncouplings}.  The region consistent with the $A^{tt}_{FB}$ measurement is shown in translucent yellow with small dots. Note that for clarity of presentation we have combined the sensitivity from  coloron plus $W$ and coloron plus $Z$ channels in this figure. 

The results are encouraging: The LHC will be able to provide information about the chiral structure for a wide range of couplings and masses and hence point us in the direction of the underlying theoretical structure. 

%%%
%%%

\section{A Flavorful Coloron Model}
In Ref. \cite{Chivukula:2013kw}, we introduced a model in which the strong interactions are extended to an $SU(3)_1 \times SU(3)_2$ structure in a way that causes the new heavy coloron states to couple differently to the third generation quarks than to the lighter generations.  What is novel here is that the model also naturally addresses the experimental observation that the third family of quarks have only a small mixing with the lighter families.  This is implemented through the presence of heavy weak-vector quarks that transform in the same way under the extended color group as the third-generation quarks. Mixing between the ordinary quark generations occurs only because all three generations mix with the vector quarks; the different gauge charges of the third and lighter generations of quarks thus leads to naturally small mixing between those generations.  Effectively, this model realizes next-to-minimal flavor violation \cite{Agashe:2005hk}.

The color gauge sector of the model remains as described above in Sec. \ref{sec:gauge}.  The matter fields of this model are summarized\footnote{The lepton fields are assigned
to $SU(2) \times U(1)$ just as in the standard model. We normalize hypercharge such
that $Q=T_3 + Y$.} below in Table \ref{table:i}. Those coupled to $SU(3)_1$ include one chiral 
quark generation ($q_L$, $t_R$, and $b_R$), which will be associated primarily with
the third generation quarks, and one vectorial quark generation ($Q_{L,R}$). The two remaining (chiral)
quark generations ($\vec{\psi}_L$, $\vec{u}_R$, and $\vec{d}_R$) are coupled to $SU(3)_2$ and
will be associated primarily with the two light quark generations. 
Noting that $Q_L$ and $q_L$ transform in the same way under the gauge symmetries, we  define $\vec{Q}_L \equiv (q_L, Q_L)$ and observe that the flavor symmetries (ignoring 
gauge anomalies) of the quark kinetic energy terms in this model are
\begin{equation}
U(2)_{\vec{\psi}_L} \times U(2)_{\vec{u}_R} \times U(2)_{\vec{d}_R} \times U(2)_{\vec{\mathcal Q}_L} \times U(1)_{t_R} \times U(1)_{b_R}
\times U(1)_{Q_R}~.
\label{eq:flavorsymmetries}
\end{equation}
As discussed in Ref. \cite{Chivukula:2013kw} the flavor symmetries may be used to simplify
our analysis of the fermion masses and Yukawa couplings.

\begin{table}
\centering
\begin{tabular}{|c|c|c|c|c|}
\hline
Particle & $SU(3)_1$ & $SU(3)_2$ & $SU(2)$ & $U(1)$ \\
\hline\hline
$\vec{\mathcal Q}_L= \begin{pmatrix}
 q_L \\
 Q_L
 \end{pmatrix}$ & 3 & 1 & 2 & +1/6\\
\hline
$t_R$ & 3 & 1 & 1 & +2/3\\
\hline
$b_R$ & 3 & 1 & 1 & -1/3\\
\hline
$Q_R$ & 3 & 1 & 2 & +1/6\\
\hline\hline
$\vec{\psi}_L = \begin{pmatrix}
 \psi^1_L \\
 \psi^2_L
 \end{pmatrix}
$ & 1 & 3 & 2 & +1/6\\
\hline
$\vec{u}_R= \begin{pmatrix}
 u^1_R \\
 u^2_R
 \end{pmatrix}$ & 1 & 3 & 1 & +2/3\\
\hline
$\vec{d}_R= \begin{pmatrix}
 d^1_R \\
 d^2_R
 \end{pmatrix}$ & 1 & 3 & 1 & -1/3\\
\hline\hline
$\phi$ & 1 & 1 & 2 & +1/2\\
\hline
$\Phi$ & 3 & $\bar{3}$ & 1 & 0 \\
\hline\hline
\end{tabular}
\caption{$SU(3)_1 \times SU(3)_2 \times SU(2) \times U(1)$ gauge charges of
the particles in this model. The $\phi$ and $\Phi$, respectively, denote the scalars responsible
for breaking the electroweak and (extended) strong sectors, while all other listed 
particles are fermions. The vectors ($\vec{\psi}_L$, $\vec{u}_R$,
$\vec{d}_R$, and $\vec{\mathcal Q}_L$) denote different fermion flavors with the
same gauge charges, where the superscripts [1,2] refer to the two light fermion generations.} \label{table:i}
\end{table}

Let us take a closer look at those mass and Yukawa terms. The existence of the  right-handed weak doublet state $Q_R$ permits the Dirac mass term
\begin{equation}
\vec{\bar{\mathcal Q
}}_L\cdot \vec{\mathcal M}\, Q_R + h.c.~,
\end{equation}
where $\vec{\mathcal M}$ is an arbitrary two-component complex mass matrix. Using the
$U(2)_{\vec{\mathcal Q}_L}$ symmetry of the quark kinetic terms, we will choose to work in a basis
in which $\vec{\mathcal M}^T=(0\ M)$ where $M$ is real and positive.  This defines what we will 
mean by $q_L$ and $Q_L$ in Table \ref{table:i} from here on.

The third-generation quark Yukawa couplings are given by
\begin{equation}
\frac{\sqrt{2}M}{v} \cdot \left(
\beta_b \bar{q}_{L} \phi  b_R + \beta_t \bar{q}_{L} \tilde{\phi} t_R
\right) + h.c.~,
\label{eq:topyukawa}
\end{equation}
where the $\beta_{t,b}$, can be chosen to be real, using the $U(1)_{t_R}\times U(1)_{b_R}$ symmetries.
The Yukawa couplings for the light two generations are given by
\begin{equation}
\frac{\sqrt{2}M}{v} \cdot \left(
\vec{\bar{\psi}}_L \phi \lambda_d \vec{d}_R + \vec{\bar{\psi}}_L \tilde{\phi} \lambda_u \vec{u}_R
\right) + h.c.~.
\end{equation}
where $\lambda_{u,d}$ are $2\times 2$ complex matrix Yukawa couplings. Neglecting the (small) mixing
of the third-generation quarks with the first two generations, the parameters $\beta_{t,b}$
and $\lambda_{u,d}$ are just equal to the corresponding parameters in the standard model,
up to the factor of $\sqrt{2}M/v$ which is included for later convenience.\footnote{While incorporating
$M$ into the Yukawa couplings is convenient for subsequent calcualtions, its presence
obscures the {\it decoupling} properties \cite{Appelquist:1974tg} of the theory in the limit
$M\to \infty$.}

Mixing of the third quark generation with the first two occurs only because all
three generations mix with the (heavy) vector quarks. Mixing between the third-generation 
quarks and the vector quarks occur through
\begin{equation}\label{eq:4-3mix}
\frac{\sqrt{2}M}{v} \cdot \left(  \lambda'_b \bar{Q}_{L}\phi b_R+\lambda'_t \bar{Q}_{L}\tilde{\phi} t_R\right) + h.c.~,
\end{equation}
and mixing between the first- and second-generation quarks and the vector quarks occurs through the Yukawa couplings to the color-octet scalar
\begin{equation}
\frac{M}{u}\cdot \left(
\vec{\bar{\psi}}_L \cdot \vec{\alpha}\, \Phi Q_R
\right) + h.c.~.
\label{eq:octetyukawa}
\end{equation}
Here the $\lambda'_{b,t}$ are complex numbers, while $\vec{\alpha}$ is a two-component complex vector,
whose phases and orientations can be simplified using the residual flavor symmetries. 
Note that in the limit that either $\lambda'_{b,t}\to 0$ or $\vec{\alpha} \to 0$,
third-generation quark number is conserved separately from
quark number for the light quarks,\footnote{In the limit $\lambda'_{t,b} \to 0$, top- and
bottom-quark number is conserved separately, while in the limit $\vec{\alpha} \to 0$ it is conserved
in combination with vector-quark number.} and mixing between the third generation and the first two vanishes.

As a result, having the mixing between the heavy and light quark generations be small is natural in this model.  In fact, we find that the pattern of quark masses and  CKM mixings is reproduced naturally in this model.  Specifically, the CKM matrix is correctly reproduced by:
\begin{align}
\begin{split}\label{est-ckm}
& V_{ub}=\alpha_1 d=\alpha_1 \left(\frac{\lambda'_t}{\beta_t}-\frac{\lambda'_b}{\beta_b}\right)=A\lambda^3(\rho-i\eta)=0.00131-i0.00334\\
& V_{cb}=\alpha_2 d=\alpha_2 \left(\frac{\lambda'_t}{\beta_t}-\frac{\lambda'_b}{\beta_b}\right)=A\lambda^2=0.0415~,
\end{split}
\end{align}
where both $d$ and $\alpha_2$ are real, and where $\alpha_2$ is ${\cal O}(\lambda^2)$ while $\alpha_1$ is ${\cal O}(\lambda^3)$.  Here, the parameters $A$, $\lambda$, $\rho$, and $\eta$ belong to the conventional Wolfenstein parameterization of the CKM matrix.

\begin{figure}
\includegraphics[width=0.8\textwidth]{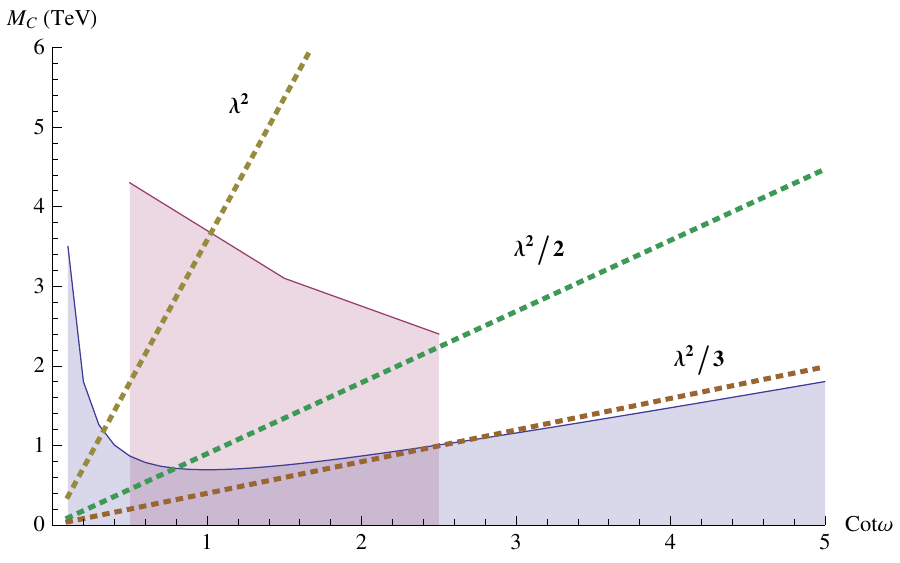}
\caption{Left:  Exclusion regions on the plane $(\cot\omega, M_C)$ from the ATLAS search for dijet resonances (pink region, beneath the short upper curve) and from $B_s$ mixing, assuming that Yukawa couplings take on the maximum value allowed by $b\to s \gamma$, (blue region, beneath the long lower curve). In addition, a bound from Kaon mixing excludes the region below the dashed line whose label matches the value of the coupling $|\alpha_2|$; a larger value yields a stronger bound.  See  Ref. \cite{Chivukula:2013kw} for details.}
\label{fig:coupling-plane-sketch}
\end{figure}

We are now prepared to examine the phenomenology of the model in more detail.

New contributions to flavor-changing neutral currents (FCNC) from the mixing of the ordinary fermions with the new weak vector femion states, and through the couplings of the colorons to fermions.  We find that data on $b \to s\gamma$ and $\Delta F = 2$ meson mixing processes place bounds on the model parameters but leave substantial regions of allowed parameter space, as indicated in Fig. \ref{fig:coupling-plane-sketch}.

We have also compared the predictions of the model to data from searches for new resonances decaying to dijets performed by the ATLAS \cite{atlas-conf-2012-148} and CMS \cite{cms-pas-exo-12-016} experiments at the LHC.  We find that the lower bound on the coloron mass ranges from $M_C \geq 2.4$ TeV for $\cot\omega \approx 2.5$, when the coloron couples mainly to third-generation quarks, all the way to $M_C \geq 4.3$ TeV for $\cos\omega \approx 0.5$, when the coloron couples more strongly to first and second generation quarks.  In contrast, the ATLAS \cite{Atlas-tt} and CMS \cite{cms-tt} searches for  resonances decaying to $t\bar{t}$ put weaker bounds on the coloron mass than the searches for resonances in dijets.  Limits from the dijet searches are also summarized in Fig.  \ref{fig:coupling-plane-sketch}.

Finally, it is worth noting that LHC data provides a lower bound on the masses of the heavy quark states that are mostly composed of the vector fermions.  The heavy $\mathsf B$, $\mathsf T$ states (partners of $b$ and $t$) can be produced at the LHC in pairs via gluon-gluon fusion \cite{Contino:2008hi, AguilarSaavedra:2009es} or singly, through their interactions with $W$, $Z$, or $h$ \cite{AguilarSaavedra:2009es, Mrazek:2009yu, Vignaroli:2012sf}. The ATLAS search in these channels \cite{Atlas-conf-130} for a 4th generation down-type quark, which decays predominantly into $Wt$, puts a limit on the 4th generation quark mass that we can directly apply to the ${\mathsf B}$ vector fermion mass: $ M_{\mathsf B}\gtrsim 0.67\ \text{TeV}$.  The analogous limit on the heavy up-type vector quark is weaker.

\section{Conclusions}

The LHC is poised to seek signs of new physics beyond the standard model in the strong-interaction gauge sector.  This paper has reviewed new developments in theory, model-building, and phenomenology that offer promising avenues for further exploration in the LHC's upcoming high-energy run.  We look forward to seeing what the experiments will discover.

\section*{Acknowledgments}

We thank John M. Campbell, Wayne W. Repko, Carl Schmidt, and Devin G.E. Walker for their input and discussions.   The authors were supported, in part, by the U.S.\ National Science Foundation under Grants No.\ PHY-0854889 and PHY-0855561.  PI is supported by Development and Promotion of Science and Technology Talents Project (DPST), Thailand.  We also acknowledge the support of the Michigan State University High Performance Computing Center and the Institute for Cyber Enabled Research.  R.S.C. and E.H.S. acknowledge the hospitality of the Galileo Galilei Institute for Theoretical Physics, where part of this work was accomplished.  A.F. acknowledges the hospitality of TASI 2011, where part of this work was completed.

\bibliographystyle{ws-procs975x65}
\bibliography{ws-pro-sample}

\begin{thebibliography}{99}
\bibitem{cdf_afb5invfb2010}
 Measurement of the inclusive forward-backward asymmetry and its rapidity dependence afb(delta y) of tt production in 5.3 fb-1 of tevatron data, CDF-note-10185.
\end{thebibliography}

\end{document}